\documentclass[]{emulateapj}

\usepackage{graphicx}
\usepackage{hyperref}
\pdfoutput=1
\newcommand{\etal}{et~al.}
\newcommand{\msun}{M$_\odot$}
\newcommand{\lsun}{L$_\odot$}

\newcommand{\ts}{\thinspace}
\newcommand{\simless}{\mathbin{\lower 3pt\hbox
     {$\rlap{\raise 5pt\hbox{$\char'074$}}\mathchar"7218$}}}
\newcommand{\simgreat}{\mathbin{\lower 3pt\hbox
     {$\rlap{\raise 5pt\hbox{$\char'076$}}\mathchar"7218$}}}

\newcommand{\about}    {$\sim$\ts}
\newcommand{\aboutless}{$\simless$\ts}
\newcommand{\aboutmore}{$\simgreat$\ts}

\slugcomment{Accepted for publication in \apj}
\shortauthors{P\'erez et al.}
\shorttitle{Atmospheric Phase Correction using CARMA-PACS: Observations of FU-Orionis star PP~13S*}
\begin{document}

\title{Atmospheric Phase Correction using CARMA-PACS: High Angular Resolution Observations of the FU-Orionis star PP~13S*}
\author{Laura M. P\'erez\altaffilmark{1}, 
James W. Lamb\altaffilmark{2}, 
David P. Woody\altaffilmark{2}, 
John M. Carpenter\altaffilmark{1}, 
B. Ashley Zauderer\altaffilmark{3}, 
Andrea Isella\altaffilmark{1}, 
Douglas C. Bock\altaffilmark{4}, 
Alberto D. Bolatto\altaffilmark{3},  
John Carlstrom\altaffilmark{5}, 
Thomas L. Culverhouse\altaffilmark{5}, 
Marshall Joy\altaffilmark{6},
Woojin Kwon\altaffilmark{7}, 
Erik M. Leitch\altaffilmark{2,5}, 
Daniel P. Marrone\altaffilmark{5,8}, 
Stephen J. Muchovej\altaffilmark{2}, 
Richard L. Plambeck\altaffilmark{9}, 
Stephen L. Scott\altaffilmark{2}, 
Peter J. Teuben\altaffilmark{3} 
and 
Melvyn C. H. Wright\altaffilmark{9}
}

\altaffiltext{1}{Department of Astronomy, California Institute of Technology, 1200 East California Blvd, Pasadena, CA 91125}
\altaffiltext{2}{Owens Valley Radio Observatory, California Institute of Technology, Big Pine, CA 93513}
\altaffiltext{3}{Department of Astronomy, University of Maryland, College Park, MD 20742-2421}
\altaffiltext{4}{Combined Array for Research in Millimeter-wave Astronomy, P.O. Box 968, Big Pine, CA 93513}
\altaffiltext{5}{Department of Astronomy and Astrophysics, University of Chicago, 5640 S. Ellis Ave. Chicago, IL 60637}
\altaffiltext{6}{Space Sciences - VP62, NASA Marshall Space Flight Center, Huntsville, AL 35812}
\altaffiltext{7}{Department of Astronomy, University of Illinois, Urbana, IL 61801}
\altaffiltext{8}{Jansky Fellow, National Radio Astronomy Observatory}
\altaffiltext{9}{Astronomy Department, University of California at Berkeley, Berkeley, CA 94720-3411}

\begin{abstract}
We present $0.15''$ resolution observations of the 227~GHz continuum emission
from the circumstellar disk around the FU-Orionis star PP~13S*. The data were
obtained with the Combined Array for Research in Millimeter-wave Astronomy
(CARMA) Paired Antenna Calibration System (C-PACS), which measures and
corrects the atmospheric delay fluctuations on the longest baselines of the
array in order to improve the sensitivity and angular resolution of the
observations. A description of the C-PACS technique and the data reduction
procedures are presented. C-PACS was applied to CARMA observations of PP~13S*,
which led to a factor of 1.6 increase in the observed peak flux of the source,
a 36\% reduction in the noise of the image, and a 52\% decrease in the
measured size of the source major axis. The calibrated complex visibilities
were fitted with a theoretical disk model to constrain the disk surface
density. The total disk mass from the best fit model corresponds to
0.06~\msun, which is larger than the median mass of a disk around a classical
T Tauri star. The disk is optically thick at a wavelength of 1.3~mm for
orbital radii less than 48~AU. At larger radii, the inferred surface density 
of the PP~13S* disk is an order of magnitude lower than that needed to 
develop a gravitational instability.
\end{abstract}


\section{Introduction}

Electromagnetic waves from an astronomical radio source suffer distortion from
irregularities in the refractivity of the atmosphere \citep{TMS2001}. At
millimeter and submillimeter wavelengths, the distortions are caused primarily
by a turbulent water vapor distribution, though dry air turbulence may also be
important under some circumstances \citep{Stirling2006}. The spatial scales of 
the turbulence extend to kilometer distances with a power law spectrum 
\citep{Kolmogorov1941} that creates atmospheric delay fluctuations on timescales that
range from fractions of a second to tens of minutes. The signal degradation is
particularly serious for millimeter-wave radio interferometers in extended
configurations, where perturbation of phases across the instrument often
exceeds many radians. These perturbations can lead to decorrelation (loss of
amplitude), distortion of the image, and loss of resolution \citep{LayI1997}. 

Different approaches have been employed to overcome the effect of atmospheric
delay fluctuations. A straightforward approach is \emph{self-calibration}
\citep{Schwab80}, where the visibility phase used to calibrate the data is
measured from the actual science target. A model for the source spatial
structure is needed and bright sources are required to measure the fringe
phase with a high signal to noise ratio in a short integration. Two other
approaches have been applied to faint targets. In \emph{water vapor radiometry}
\citep{Westwater1967, Woody2000}, a dedicated radiometer monitors the water
vapor emission along the pointing direction of the antenna. In \emph{fast
position switching} \citep{Holdaway1995}, the antennas switch rapidly
between the science target and a nearby phase calibrator to capture the
atmospheric delay fluctuations on time scales longer than the switching cycle
time. Both methods probe the atmosphere close to the line-of-sight toward the
science target, but have limitations in actual applications. Fast position
switching reduces the time spent on source by a factor of $\sim 2$, while water
vapor radiometry requires a physical model to relate the water line brightness
and the path correction, as well as it assumes that refractivity fluctuations are
produced only by water vapor.

As an alternative approach, an array of closely \emph{paired antennas}
\citep{Asaki1996, Asaki1998} continuously monitors the atmospheric phase
fluctuations by observing a nearby calibrator. Two arrays of antennas are
necessary: antennas belonging to the ``science array" observe the science
target and phase calibrator, while antennas belonging to the ``reference array"
simultaneously monitor an atmospheric calibrator. Phase correction on the
science target and phase calibrator is accomplished by subtracting the
visibility phase measured from the atmospheric calibrator on each baseline. An
advantage of the paired antenna technique over water vapor radiometry is that
it accounts for atmospheric phase fluctuations due to both water vapor and a
dry air component.

Between November 2008 and February 2009, the Combined Array for Research in
Millimeter-wave Astronomy (CARMA) implemented the CARMA Paired Antenna
Calibration System (C-PACS) to correct for atmospheric delay fluctuations on
the longest baselines of the interferometer (up to 2~km in length). The goal of
C-PACS is to enable routine imaging in the most extended CARMA configurations.

In this paper we describe C-PACS and apply this calibration technique to
observations of the circumstellar dust around PP~13S*\footnote{PP~13 is a
cometary nebula in the list of \citet{Parsamian1979}. PP~13S is the southern 
component containing a red nebula with a bright infrared point source at the 
apex as designated by \citet{Cohen1983}. PP~13S* corresponds to the embedded
star itself, which is obscured by circumstellar material. The northern 
component, PP~13N, is a T~Tauri star.}. PP~13S* is a young pre-main sequence
star located in the constellation of Perseus and embedded in the L1473 dark
cloud at a distance of $\sim$ 350 pc \citep{Cohen1983}. This object is thought
to have experienced a FU-Orionis-type outburst in the past due to a massive
accretion episode and is now declining in brightness to a quiescent state
\citep{Aspin2001}. The FU-Orionis nature of PP~13S* has been established based
on the highly broadened infrared CO absorption bands \citep{Sandell1998}, the
jet-like feature seen in [SII] emission which is characteristic of Herbig-Haro
outflows \citep{Aspin2000,Aspin2001}, and the consistent dimming and morphology
changes observed at near-IR and optical wavelengths over several decades
\citep{Aspin2001}. All of these characteristics are common to FU-Orionis
objects \citep{Hartmann1996}. The star, with a bolometric luminosity of
30~\lsun\ \citep{Cohen1983}, is surrounded by an extended disk and envelope
that contains about 0.6~\msun\ of gas and dust \citep{Sandell1998}. The new
CARMA observations will help understand the origin of the FU-Orionis
phenomena in PP~13S* by measuring the distribution of dust continuum emission
on spatial scales of $\sim 50$ AU.

\section{Description of C-PACS}
\label{sec:cpacs}

Before presenting the new observations of PP~13S*, we describe the paired
antenna calibration system as implemented at CARMA. We first describe the
antenna configuration used for the observations, and outline the basic
principles of the technique.

\subsection{CARMA}
\label{sec:carma}

CARMA is an heterogeneous interferometer comprising 23 antennas: six 10.4-m 
telescopes from the California Institute of Technology/Owens Valley Radio 
Observatory (OVRO), nine 6.1~m telescopes from the Berkeley-Illinois-Maryland 
Association (BIMA), and eight 3.5-m telescopes from the University of Chicago
Sunyaev-Zel'dovich Array (SZA). The two most extended array configurations
contain baselines that range in length from 100~m to 1000~m (B configuration)
and 250~m to 1900~m (A configuration) to achieve an angular resolution of
0.3{\arcsec} and 0.15{\arcsec} respectively, at an observing frequency of
230~GHz.

A schematic of C-PACS is shown in Figure \ref{C-PACS}. C-PACS pairs the eight
3.5-m antennas with selected 6-m and 10-m antennas, preferentially on the
longer baselines. Typically, a 3.5-m antenna is offset by 20--30~m to the west
of a larger antenna. This separation is a compromise between the need to
put the antennas as close as possible to probe the same atmospheric path and
to avoid shadowing between antennas. 

The science array, composed of the 6-m and 10-m antennas, operates in the 1.3-
or 3-mm atmospheric windows as requested by the investigator. The reference
array, comprising the 3.5-m antennas, operates in the 1-cm window. The
observing cycle consists of observations of the science target interleaved with
periodic observations of the phase calibrator. Both the science and reference
arrays observe the phase calibrator to measure instrumental phases drifts.
Subsequently, the science array observes the science target while the reference
array monitors a strong point source (i.e., the ``atmospheric'' calibrator)
close to the science target, to measure the delay introduced by the atmosphere.
The atmospheric delay measured by the 3.5-m antennas can then be applied
to the science observations. 

\subsection{Properties of the atmosphere}
\label{sec:atmos}

It is helpful to have a physical picture to understand the principles of the
correction. We suppose the atmosphere to be a pattern of random
refractive index variations that is blown across the array at the wind
velocity. Furthermore, we assume that the layer is at a height of a couple of
kilometers and that the thickness is much less than the height. General
experience at this and other sites show that these conditions are often
consistent with what is observed. Other conditions can be present, but can
often be characterized by two or more layers at different altitudes with
separate wind vectors so that only a small change to the analysis is required.

Kolmogorov theory predicts a turbulence distribution 
with a power law spectrum from less than a millimeter in size to many
kilometers. This results in random delay differences between signals arriving
at different antennas that increase with separation as a power law function
\citep{Tatarski1961}. As the pattern moves over the array, the delay
differences are observed as temporal fluctuations in the visibility phases. The
RMS of the delay depends on the wind speed, but not its direction. 
Structures smaller in size than the antenna diameter are averaged out and do not
contribute to the phase errors. Structures on scales large in comparison
with the baseline length are common to the two antennas on the baseline and
therefore tend to cancel out. From these theoretical considerations supported
by experimental evidence \citep{Sramek83,Sramek89}, it is found that the 
resulting delay variance vs baseline length (i.e., the delay structure 
function) also follows a power-law. The theoretical slope of the power-law 
is 5/3 and 2/3 for two- and three-dimensional Kolmogorov turbulence, 
respectively.

\subsection{Atmospheric delay corrections}

\begin{figure}
\begin{center}
\includegraphics[scale=1.5]{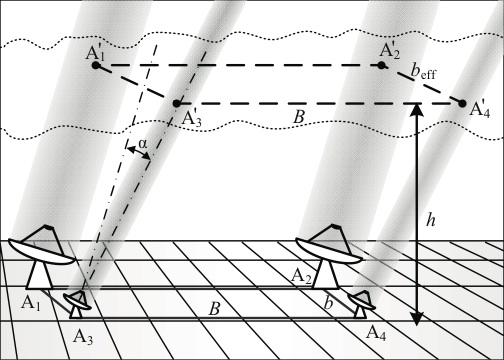}
\caption{Schematic of the CARMA Paired Antenna Calibration System. Antennas $A_1$ and $A_2$ (in the science array) observe the science target while antennas $A_3$ and $A_4$ (in the reference array), offset by a distance $b$ from the science array, observe a bright atmospheric calibrator. In the turbulent layer at an elevation $h$ the beams are separated by a distance $b_{\mathrm{eff}}$ that depends on $b$, $h$, and the angular offset between the science and reference sources, $\alpha$. If the effective baseline, $b_{\mathrm{eff}}$, is much smaller than the science baseline, $B$, the delay difference to the reference antennas is a good estimate for the delay difference to the science antennas.}
\label{C-PACS}
\end{center}
\end{figure}

Following the discussion in \citet{Asaki1996}, consider two pairs of antennas
as shown in Figure \ref{C-PACS}. In the simplest case, we measure the delay
difference between the antennas on the reference baseline (i.e., baseline
$A_3-A_4$ in Figure~\ref{C-PACS}) as a function of time. Assuming a non-dispersive
atmosphere, the delay on the science baseline (i.e., baseline $A_1-A_2$) is
corrected by applying the delay difference on the reference baseline to the
visibility measurement.

\begin{figure}
\begin{center}
\includegraphics[scale=1.0]{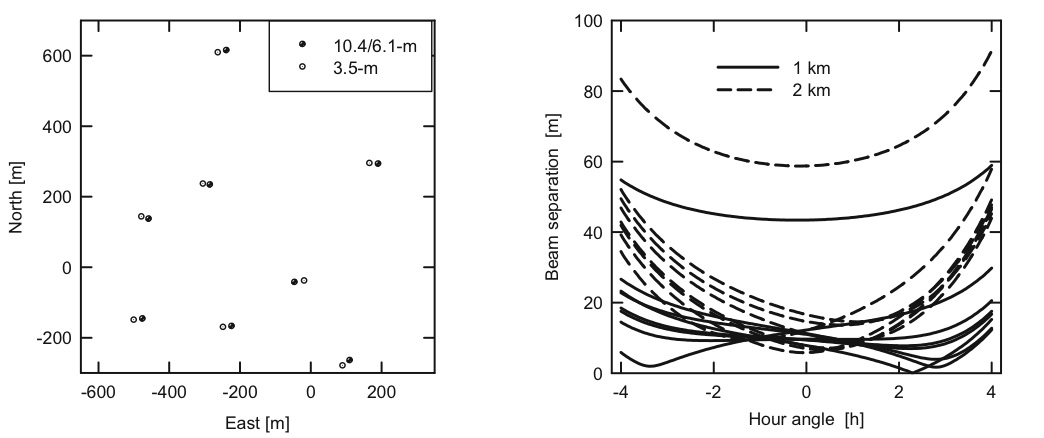}
\caption{The left panel shows the locations of the science (10.1-m and 6.4-m) and reference (3.5-m) antennas in the B-configuration of CARMA (circles do not represent the actual antenna diameters). The right panel shows the separation, $b_{\mathrm{eff}}$, between the science and reference beams in the turbulent layer for each pair of antennas during the PP~13S* observations. The continuous (dashed) line corresponds to a turbulent layer at a 1(2)-km altitude. For most pairs the configuration is very favorable, giving rise to effective baselines much shorter than the science baseline.}
\label{fig:ant} 
\end{center} 
\end{figure}

The reference and science delays are not identical since the two baselines are
not exactly co-located. The relevant distances that determine the efficacy of
the delay corrections are not the baseline lengths at the ground, but the
distances between the radio beams as they traverse the turbulent layer (i.e. $A_1' - A_3'$ and $A_2' - A_4'$ in Figure \ref{C-PACS}). The
beam separation at the turbulent layer depends upon the relative positions of
the target and reference source in the sky, the height of the turbulent layer,
and the configuration of the antennas on the ground. The upper limit to the
beam separation is given by 
\begin{equation} 
   \label{eq:dmax}
   d_\mathrm{max} = |A_1-A_2| + \alpha h/sin(e), 
\end{equation} 
where $\alpha$ is the angular separation between the science target and the
atmospheric calibrator, $h$ the height of the turbulence, and $e$ is the
source elevation. Assuming that the turbulent layer is at a height of 1~km 
(continuous line) or at 2~km (dashed line),
Figure~\ref{fig:ant} shows the trajectories of the 3.5-m beam locations
relative to the 6-m and 10-m beams for an 8~h observation of PP~13S* centered
on transit with 3C111 as the atmospheric calibrator. As shown in this figure,
the choice of 3C111 as the atmospheric calibrator is particularly fortuitous 
for PP~13S* since the science and reference beams for most antennas nearly 
cross within the turbulent zone.

Using phase closure \citep{Jennison58}, the difference ($\Delta\tau$) between
the actual delay for the target beams $A_1'-A_2'$ and the atmospheric reference
beams $A_3'-A_4'$ is given by $\Delta\tau = \tau(A_1'-A_2') - \tau(A_3'-A_4') =
\tau(A_1'-A_3') - \tau(A_2'-A_4')$. In a favorable configuration, the beam
separations $A_1'-A_3'$ and $A_2'-A_4'$ will be much less than either the
target beam separation $A_1'-A_2'$ or the reference beam separation
$A_3'-A_4'$. The RMS of the corrected visibilities will be $\sqrt{2}$ worse
than for an array with beam separation $A_1'-A_3'$, which implies that C-PACS
will have a performance equivalent to an array that has baseline lengths
{$\sim$}20--70{\%} larger than the $A_1'-A_3'$ beam separation, depending on
the structure function exponent. The complete analysis contains additional
correlation terms and added uncertainties caused by the finite signal-to-noise
for the atmospheric calibrator observations.

\subsection{Atmospheric calibrators}
\label{calibrators}

The ability of C-PACS to correct the atmospheric delays is limited by
the delays from the short beam spacings $A_1'-A_3'$ and $A_2'-A_4'$ (see 
Figure~\ref{C-PACS}), instrumental phase drifts on the reference array, and 
the radiometer noise. The delay errors caused by differences in the beam 
spacings between the science and reference arrays are given statistically by 
the structure function $R(|A_i'-A_j'|)$. The instrumental errors can be removed 
by removing a box-car average over the length of the observation, and 
will contribute negligible delay errors as long as the timescale for the 
instrumental drifts are large compared to the box-car width. 
The delay errors due to radiometric phase noise depend on the strength of 
the source being observed, the receiver properties, and the atmospheric 
characteristics \citep{TMS2001}. The uncertainty in the measured phase from 
radiometer noise is given by
\begin{eqnarray}
\Delta \phi = \frac{\sqrt{2}k_{\mathrm{B}}T_{\mathrm{sys}}}{\eta_Q A_{\mathrm{eff}} S 
\sqrt{B t}}
\label{radiometric_noise}
\end{eqnarray}
where $k_{\mathrm{B}}$ is the Boltzmann's constant, $T_\mathrm{sys}$ is the 
system temperature, $\eta_Q$ is the correlator quantization efficiency, $A_\mathrm{eff}$ is
the effective collecting area of the antennas, $S$ is the flux density of the atmospheric 
calibrator, $B$ is the bandwidth of the observations, and $t$ is the integration time.
The measurement uncertainty in the delay is then $\Delta \phi / (2\pi\nu)$. The net variance in the delay after applying C-PACS is given by 
\begin{equation}
  \label{eq:dtau}
  \Delta\tau^2 \approx R(|A_1'-A_3'|) + R(|A_2'-A_4'|) + 
        \Bigl({k_B T_\mathrm{sys} \over \eta_Q A_\mathrm{eff} S \pi \nu \sqrt{2 B t}}\Bigr)^2 
\end{equation}
The structure function is generally described by a power law with exponents
varying from 5/3 to 2/3 depending upon the spacing and thickness of the
turbulent layer. The scaling coefficient of the power law also varies depending
upon the weather conditions. In order for C-PACS to improve the image quality,
the target and reference beams must be close at the turbulent layer such that
$R(|A_1'-A_3'|) + R(|A_2'-A_4'|) \ll \lambda^2$. This requires angular separations
\aboutless 5\arcdeg\ for the A and B configuration C-PACS pairings and typical
winter weather conditions at the CARMA site. A future publication will
use actual measurements to quantify how the quality of the C-PACS correction
varies with angular separation between the science target and the atmospheric
calibrator (Zauderer \etal, in preparation).

The radiometer noise should also contribute much less than a wavelength of delay 
error for the C-PACS corrections to be successful. For the characteristics of 
the 3.5-m telescopes and the 1-cm receivers, the radiometer delay error is 
given by 
\begin{equation}
\label{eq:delayerror}
\Delta\tau_\mathrm{radiometer} = 1.3 \mathrm{mm} 
      \,\Bigl({S_\nu \over \mathrm{1 Jy}}\Bigr)^{-1}
      \,\Bigl({t \over \mathrm{1 s}}\Bigr)^{-1/2}
\end{equation}
Thus, 1.3-mm observations with integration times of $t=4$~s (short enough to
measure and correct most of the atmosphere fluctuations) require a reference
source brighter than $S\sim1$~Jy in the 1~cm band. When several atmospheric calibrators are
available, the optimum choice between calibrator separation and brightness
can be found by minimizing Equation~\ref{eq:dtau} for the expected weather 
conditions.

We combined the SZA 30~GHz calibrator list, the GBT catalog at 1.4 and 5.0~GHz
\citep{Condon2001}, and the WMAP point source catalog \citep{Wright2009} to
estimate the density of potential C-PACS calibrators. For each source in the
GBT catalog, we extrapolated the flux density from 5.0~GHz to 30~GHz by
measuring the spectral index $\alpha$ between 1.4 and 5.0~GHz ($S_{\nu} \propto
\nu^{\alpha}$). We find that $50 \%$ of the sky is within $5\degr$ of a point
source with flux density greater than $1$~Jy at 30~GHz. The number of suitable
C-PACS calibrators could be expanded by increasing the sensitivity of the
reference array, which would allow us to employ fainter atmospheric
calibrators. This could be accomplished by increasing the correlator 
bandwidth or improving the receiver sensitivity.

\newpage
\section{Observations and Data Reduction}
\label{obs}

PP~13S* is particularly well suited for C-PACS observations since the nearest
atmospheric calibrator (3C111) is bright (\about 4~Jy at 1.3~mm at the time of
the observations) and separated by 1.5\arcdeg\ from PP~13S*. Thus the
calibrator satisfies the basic criteria needed for successful C-PACS
corrections (see Section~\ref{calibrators}). In this section, we describe the
CARMA observations and data reduction of PP~13S*.

\subsection{CARMA 1.3-mm wavelength observations}
The 6-m and 10-m antennas were used to obtain 1.3~mm continuum observations of
PP~13S* on UT Dec 5, 2008 in the CARMA B configuration and on UT Jan 18, 2009
in the CARMA A configuration. Double-sideband receivers mounted on each antenna
were tuned to a rest frequency of 227~GHz placed in the upper sideband. The
correlator was configured with three 468.75~MHz wide bands to provide 1.41~GHz
of continuum bandwidth per sideband. The observing sequence interleaved 3~min
observations of 3C111 with 12~min observations of PP~13S* in B configuration
and 4~min in A configuration. The complex visibilities were recorded every 4~s.

Data reduction was performed using the Multichannel Image Reconstruction, Image
Analysis and Display (MIRIAD) software package \citep{Sault1995}. Each night of
observations was calibrated separately. The calibration consisted of first 
applying a line-length correction\footnote{The CARMA line-length system
measures the total round trip delay caused by possible mechanical effects and
temperature variations of the fiber-optic cables running to each antenna.} and
then a passband correction derived from observations of 3C111. Only these two
calibrations were applied to the millimeter data before proceeding with the
C-PACS corrections (see Section \ref{applying_C-PACS}). All images that are
presented here were formed by inverting the visibility data using natural
weighting, and then ``cleaning'' with the point spread function (i.e. the 
``dirty'' beam) using a hybrid H\"ogbom/Clark/Steer algorithm 
\citep{Hogbom74,Clark80,Steer84}.

\subsection{CARMA 1-cm wavelength observations}
\label{sec:cm}

The eight 3.5-m antennas were used to obtain 1-cm 
observations of the atmospheric calibrator simultaneously with the 1.3-mm
wavelength observations. Single-sideband receivers mounted on each antenna were
tuned to a sky center frequency of 30.4~GHz. For these antennas, a wideband
correlator is available that was configured with fourteen 500~MHz wide bands
to provide 7~GHz of continuum bandwidth. Complex visibilities were recorded
every 4~s in order to track the rapidly varying atmospheric fluctuations. 
Single-sideband system temperatures for the 1-cm observations ranged between 35
and 55~K. 

The data calibration consisted of applying a time-dependent passband measured from
an electronically correlated noise source that was observed every 60~s. This
passband was computed and applied on a 60~s timescale to remove any delay
variations in the digitizers due to temperature cycling of the air-conditioning
that cools the correlator. A passband correction derived from observations of
3C111 was then applied.

\subsection{Applying 1-cm delays to the 1.3-mm data}
\label{applying_C-PACS}
Phase referencing would normally be performed at this point in the data
reduction process to remove the slowly varying phase drift introduced by the
instrument. However, a phase calibration computed over a long time interval,
and prior to correcting for the ``fast'' atmospheric delay fluctuations will
alias the fast component into a slowly varying error on the phase calibration
\citep{LayII1997}. C-PACS can reduce the errors introduced from standard phase
calibration techniques by correcting for both fast and slow atmospheric
delay fluctuations. 

The delays were extracted from the 1-cm wavelength observations of the
atmospheric calibrator, and then applied to each corresponding paired baseline
in the science array. Only eight of the 15 antennas from the science array,
those paired with a reference antenna, have the C-PACS correction applied.
The delay derived from the reference array is applied to the science target
(PP~13S*) and phase calibrator (3C111) on each record. The delays were
computed from the mean phase across all channels in the 1-cm observations
divided by the mean frequency. The observed wavelength of 1~cm used for the
reference array is longer than the typical atmospheric delay fluctuations, and
delay tracking through phase wraps was not a problem. 

The atmospheric delays derived from the 1-cm data were applied directly to the
1.3-mm data without any corrections for differences in the observed frequency.
This is possible since the dispersion in refractivity of water vapor between
centimeter and millimeter wavelengths is less than a few percent
\citep{Hill1988} away from the strong atmospheric emission lines, and
ionospheric effects are negligible at these frequencies \citep{Hales2003}. The
minimal dispersion of the refractivity has also been verified experimentally
from the C-PACS observations (see Section~\ref{3C111_results}). 
After applying the delay corrections to the 1.3~mm data on 4~s intervals, a 
long-interval (10~min) phase calibration is applied to the 1.3-mm data to
remove the slow varying instrumental delay difference from the two different
arrays. 

\section{Application of C-PACS}

The effect of the C-PACS on the calibrated phases can be analyzed in several
stages. We first compare the phase fluctuations measured toward 3C111 at
wavelengths of 1~cm and 1.3~mm to demonstrate that the reference and science
arrays are tracing the same atmospheric fluctuations. We then demonstrate
that C-PACS yields quantitative improvement in the quality of the PP~13S* star
image. Throughout this section, no absolute flux scale or amplitude calibration
are applied to the data in order to evaluate how C-PACS improves the phase
stability on the phase calibrator (Section~\ref{3C111_results}) and PP~13S*
(Section~\ref{PP13S_results}).

\subsection{3C111} 
\label{3C111_results}
\begin{figure}
\begin{center}
\includegraphics[scale=0.7]{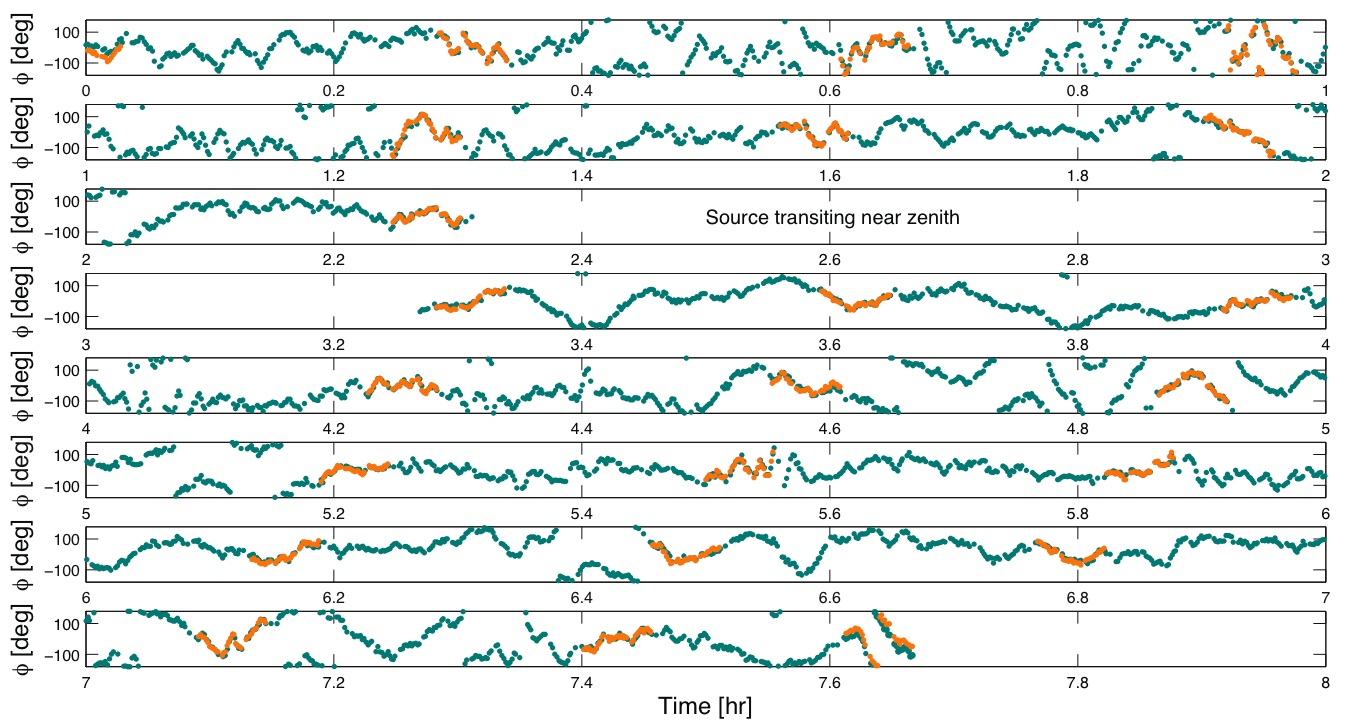}
\caption{Visibility phase versus time for the science array (orange) and
the reference array (green) measured toward 3C111 on a baseline of
length $\sim800$~m in B configuration. The phase measured by the reference
array at a frequency of 30.4~GHz was scaled to the frequency of the science
array (227~GHz) by a factor of 7.5 (= 227~GHz / 30.4~GHz). Each
data point indicates the measured phase over a 4~s integration after removing
the mean phase computed in 10~min intervals. About 1~h of data are omitted near
the middle of the observation when PP~13S* and 3C111 were transiting at an
elevation $> 80 \degr$ and the tracking of the antennas was poor. The figure
demonstrates that the observed phases at 30.4~GHz closely track the phases
measured at 227~GHz and can be used to correct the atmospheric fluctuations
at higher frequencies.
} 
\label{single_baseline} 
\end{center} 
\end{figure}

Figure \ref{single_baseline} shows the measured fringe phase toward 3C111 for
one paired baseline in the B configuration to illustrate the correlation that
exists between the phases measured at wavelengths of 1.3 mm and 1 cm. For this
figure, the phases measured on the reference array were scaled by the
ratio of the observed frequencies (227~GHz / 30.4~GHz = 7.5). The correlation
between the 1-cm and 1.3-mm phases is evident over the nearly 8~h time period
and is present for all paired baselines. In addition, the phases are tracked
between the two arrays even though the science array is switching between
two sources. These results demonstrate that (1) the 1-cm phases can be used to
track the delay fluctuations at higher frequencies, (2) the atmosphere is
non-dispersive at these wavelengths such that a linear scale with frequency can
be used to predict the phases fluctuations at other wavelengths, and (3)
C-PACS can correct the science observations while preserving the link to the
phase calibrator observations in the science array.

\begin{figure}
\begin{center}
\includegraphics[scale=0.8]{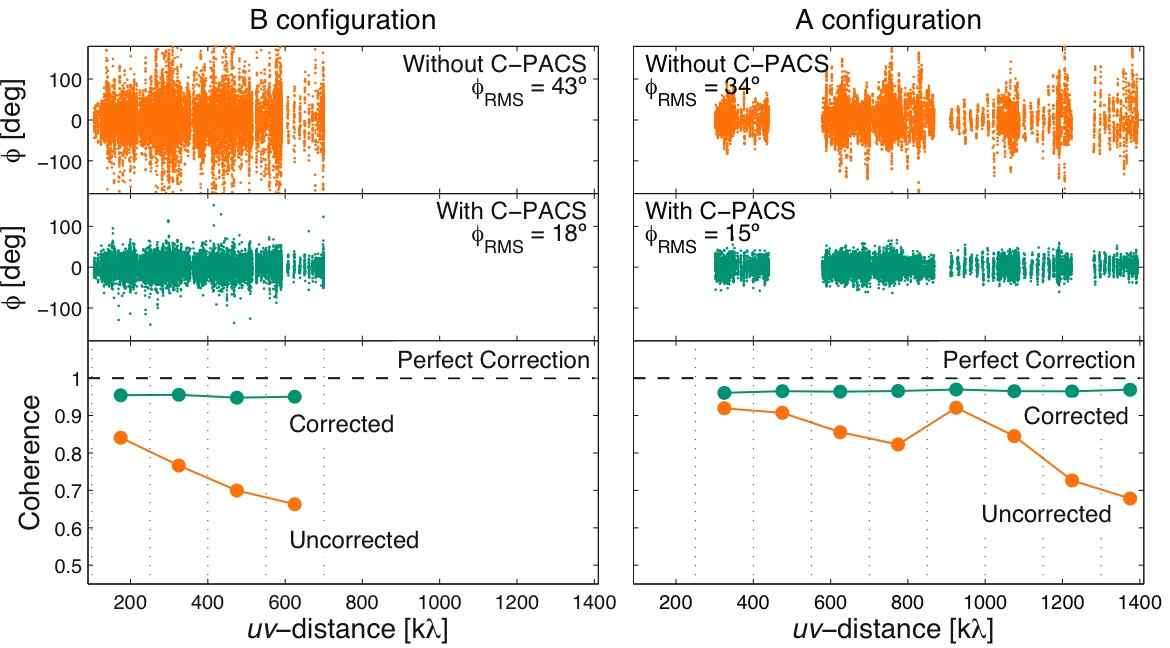} 
\caption{Visibility phases versus $uv$-distance on all paired baselines for 
observations of 3C111 in the CARMA B and A configurations. The phases are
shown before (top panels) and after (middle panels) applying C-PACS 
corrections. Each point indicates the measured phase in a 4~s integration after
removing the mean phase computed in 10~min intervals. The bottom panels show
the coherence calculated over 130~k$\lambda$ intervals. Before applying 
C-PACS, the coherence declines with $uv$-distance as expected for
atmospheric phase fluctuations that increase with baseline length. After 
applying C-PACS, the phase coherence is higher and uniform with baseline
length.
} 
\label{all_data}
\end{center} 
\end{figure}

The observed phases at 227~GHz toward 3C111 on all paired baselines in the A
and B configurations are shown as a function of $uv$-distance\footnote{The
$uv$-distance is the baseline distance projected perpendicular to the line of
sight.} in Figure~\ref{all_data}. The upper panels show the visibility phases
measured in 4~s integrations before applying C-PACS corrections, and the middle
panels show the visibility  phases after applying the corrections. The RMS scatter on the
visibility phase before correction is between 33\arcdeg\ (at short 
$uv$-distances) and 53\arcdeg\ (at long $uv$-distances) for B configuration
and between 26\arcdeg\ and 48\arcdeg\ for A configuration. After
applying the C-PACS correction, the phase scatter is reduced to $15-18\degr$ 
for correction across all $uv$-distances in both configurations.

Another way to grasp the effect of the C-PACS corrections can be seen on the
bottom panels in Figure~\ref{all_data}, where the coherence value
($e^{-\phi_{\rm{RMS}}^2/2}$) is measured over $uv$-distance bins of width
130~k$\lambda$. Before applying the C-PACS corrections, the coherence 
decreases with increasing baseline length since the longer baselines have
larger atmospheric fluctuations. After applying the C-PACS corrections,
the coherence is uniform with baseline length at a value of \about 95\%.
The coherence becomes nearly constant with baseline length
since C-PACS converts the 200-1800~m baselines into \about 30-50~m effective
baselines for all paired antennas. While 
the results shown in Figure~\ref{all_data} emphasize the improvement in
coherence, C-PACS also improves the visibility phases, which results in higher image fidelity.

\subsection{PP~13S*} 
\label{PP13S_results}

\begin{figure}
\begin{center}
\includegraphics[scale=1.0, viewport = -100 0 2124 500]{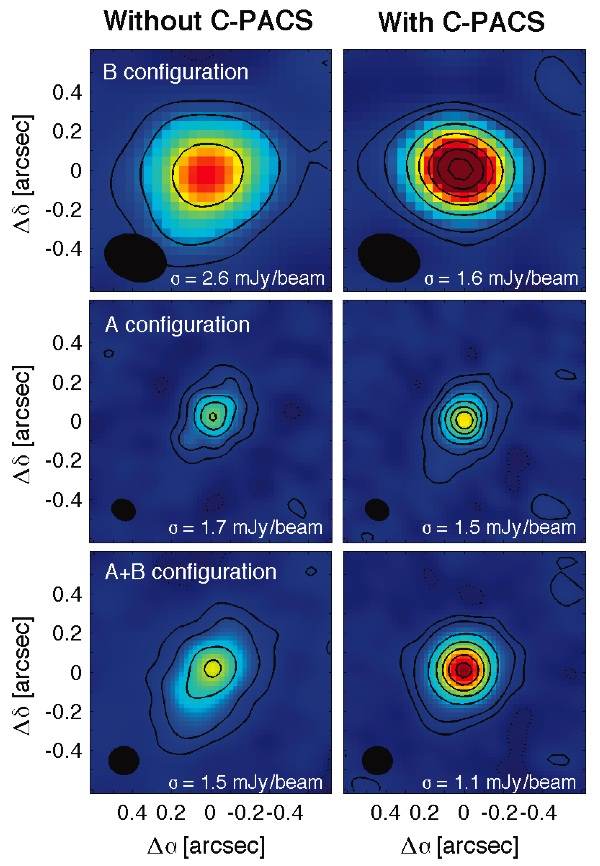} 
\caption{CARMA 227~GHz continuum images of PP~13S* before (left) and after (right)
C-PACS corrections for data obtained in B configuration (top), A configuration 
(middle), and B and A configurations combined (bottom). No absolute flux scale 
or amplitude gain calibration have been applied to the data in
order to assess the effect of atmospheric phase corrections only.
The color scale range is same for all maps (from -2.4 to 54.8~mJy) such that 
the measured fluxes can be compared directly. Solid contours are at 
2$\sigma$, $5\sigma$ and in increments of 5$\sigma$ thereafter. Dotted contour
is at $-2\sigma$. For both B and A configurations, applying the C-PACS 
correction increased the observed peak flux and reduced the observed source 
size.
}
\label{pp13s} 
\end{center} 
\end{figure}

After applying the C-PACS atmospheric delay corrections to the 1.3-mm data, 
we removed the instrumental delay drifts by phase referencing to the 3C111 
observations. 
Figure \ref{pp13s} shows the resulting maps before and after applying the
C-PACS corrections for the A and B configurations separately and the 
combined data sets.

\begin{figure}
\begin{center}
\includegraphics[scale=0.5, viewport = -10 0 2124 550]{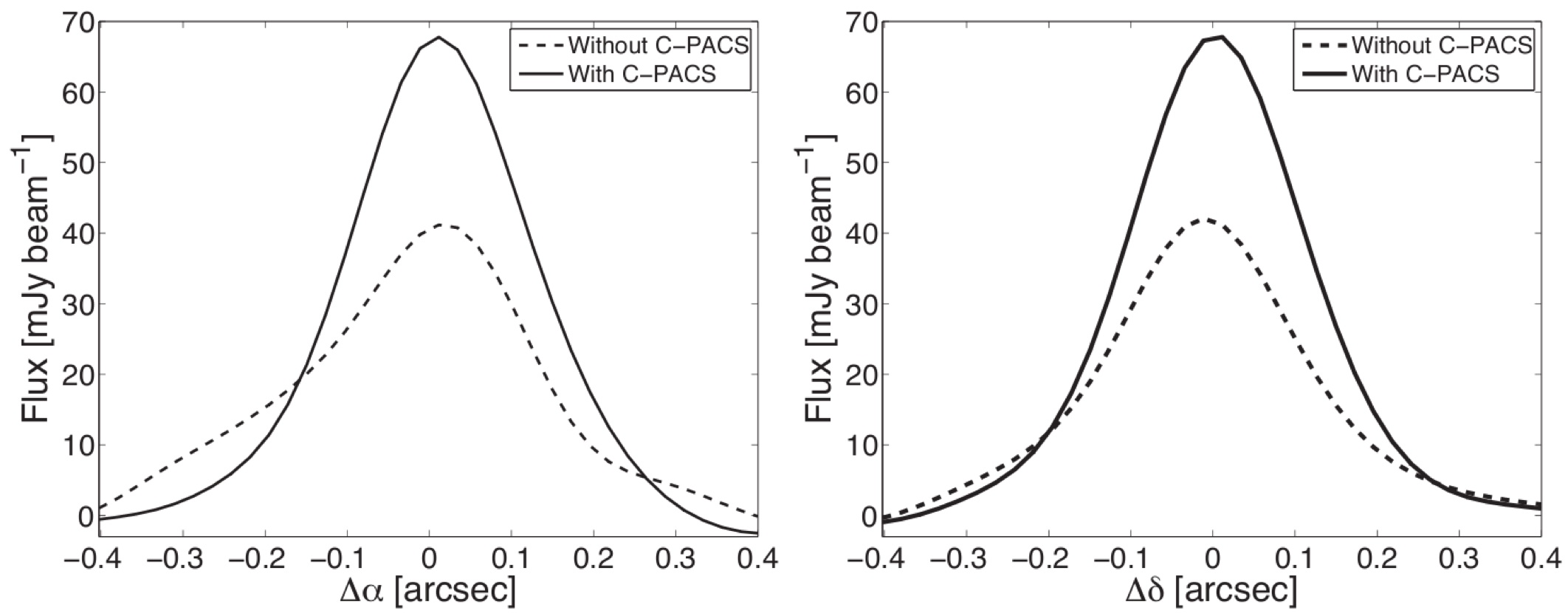} 
\caption{Observed flux density toward PP~13S* versus offset in right 
ascension (left) and declination (right) before and after 
applying C-PACS corrections. Applying C-PACS corrections increased
the peak flux density and decrease the observed source size.
}
\label{radial_profile} 
\end{center} 
\end{figure}

C-PACS improved the image quality for both the A and B configuration maps as
measured by the increase in the peak flux, the reduction in the noise level,
and the decrease in the observed source size. The improved image quality 
resulted from correcting the phase fluctuations in the 1.3~mm data. 
The increase in the source flux and decrease in source size is also illustrated in
Figure~\ref{radial_profile}, which shows radial profile plots across the
1.3~mm emission toward PP~13S* along right ascension and declination.
In the combined A+B configuration map, the peak flux measured
toward PP~13S* increased from 42.4~mJy to 67.8~mJy (a factor of 1.6) after
applying the C-PACS correction. The noise level decreased from $\sigma =
1.5$~mJy/beam to $\sigma = 1.1$~mJy/beam, which corresponds to a 36\%
improvement. The observed full-width-at-half-maximum (FWHM) source size
diminished from $0.41\arcsec\times0.27\arcsec$ to
$0.27\arcsec\times0.26\arcsec$, or a 52\% decrease in the size of the major
axis of the 1.3~mm emission. 

\section{Properties of the PP~13S* Circumstellar Disk}

Before proceeding to the analyze the properties of the dust surrounding
PP~13S*, we must flux calibrate the C-PACS data. The absolute flux calibration
was set from observations of Uranus in B configuration and 3C84 in A
configuration. The flux density of Uranus was inferred from a planet model,
while the flux density of 3C84 was obtained from CARMA observations on a
different day when both 3C84 and Uranus were observed. The uncertainty on this
calibration is estimated to be 20\% due to uncertainties in the planetary model
and the bootstrapped flux for 3C84. The antenna gains as a function of time
were then determined from the 3C111 observations. 

\begin{figure}
\begin{center}
\includegraphics[scale=0.9]{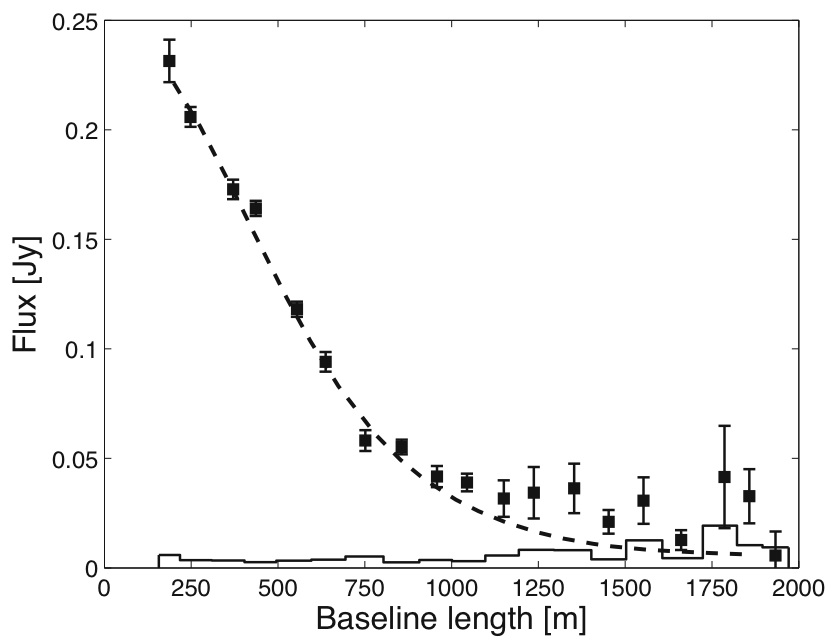} 
\caption{Observed visibility amplitudes toward PP~13S* after 
applying the C-PACS corrections (black points) versus the deprojected 
baseline length. The errorbars indicate the 1$\sigma$ interval uncertainties,
but exclude uncertainties in the flux calibration (\about 20\%). The 
histogram shows the expected signal for random noise. The dashed line 
shows the best-fit disk model (see text) to the visibility data. The
clear decline in the visibility amplitude with baseline length indicates
the dust emission around PP~13S* has been resolved.
} 
\label{uvamp} 
\end{center} 
\end{figure}

Figure \ref{uvamp} shows the calibrated visibility amplitude observed toward
PP~13S* as a function of baseline length. An unresolved source will have a
constant flux density with baseline length. By contrast, the visibility
amplitude toward PP~13S* decreases with increasing baseline length, which
suggests that the source is resolved. While the decline in amplitude with
baseline length could be explained by a coherence as low as $\sim 0.2$ on
\aboutmore 1-km baselines, the minimum measured coherence on 3C111 at 
any $uv$-distance was 0.65 even before applying the C-PACS correction (see
Figure \ref{all_data}). A similar decline in the visibility amplitude for
PP~13S* with baseline length can be seen as well when only a fraction of the
data are averaged together (for example using 1~h of data at a time),
indicating that atmospheric decorrelation over long timescales is not giving
rise to the amplitude drop at long baselines. Thus, the primary cause of the
decrease in amplitudes with increasing baseline length is that the source is
resolved.

The FWHM of the 1.3~mm continuum emission toward PP~13S* is
$0.22\arcsec\times0.21\arcsec$, which was obtained by fitting a two-dimensional
gaussian to the surface brightness distribution in the combined A+B
configuration image after applying the C-PACS corrections and deconvolving the
synthesized FWHM beam size of $0.15'' \times 0.14''$. The integrated flux
density of PP~13S* in the C-PACS corrected image after flux and amplitude
calibration is $241 \pm 48$~mJy, measured by integrating over an aperture of
radius $R_{disk} \sim 1.4 \times \mathrm{FWHM} \sim 0.36''$, where $R_{disk}$
is defined to encompass 95\% of the emission. The flux observed with CARMA
corresponds to about half of the emission measured by single-dish observations
\citep[450~mJy at 1.3~mm with a beam size of $19.5''$][]{Sandell1998}. The
remaining 1.3~mm flux is presumably contained in an extended envelope larger
than $1.6''$, which is the largest angular scale probed by the CARMA data. 

The presence of a circumstellar disk in PP~13S* has been previously inferred
from several lines of evidence: (i) reflected light along the outflow axis is
observed despite the fact that the central object is heavily obscured in the
optical \citep[$A_{V} \sim 30-50$;][]{Cohen1983}, which indicates that the
circumstellar material is not spherically symmetric; (ii) infrared absorption
bands at 3\micron\ and 10\micron\ indicate substantial
quantities of cold dust, probably present in an obscured inclined disk
\citep{Cohen1983,Smith1993}; and (iii) the broad $2.2~\mu$m CO overtone
absorption feature present in the PP~13S* spectra \citep{Sandell1998,Aspin2001}
can be explained by the presence of a massive accreting circumstellar disk
\citep{Hartmann1996}. Nonetheless, without observations of the gas kinematics,
we cannot determine if the dust emission detected by CARMA originates from the
central cusp of an envelope or from the circumstellar disk that surrounds
PP~13S*. Since the presence of a massive accretion disk has been invoked to
explain several characteristics of FU-Orionis objects, we assume that the
millimeter continuum emission around PP~13S* observed by CARMA originates
primarily from a circumstellar disk. 

To determine the disk properties, we assumed the radial surface density 
[$\Sigma(R)$] can be described by the similarity solution for a viscous
accretion disk given by
\begin{eqnarray} 
   \Sigma(R,t) = \Sigma_t \left( \frac{R_t}{R} \right)^{\gamma} \exp 
      \left[ -\frac{1}{2(2-\gamma)} 
      \left[ \left( \frac{R}{R_t} \right)^{2-\gamma} -1
      \right] \right], \;\;\;\;
\label{eq:sd}
\end{eqnarray} 
where $\gamma$ is the slope of the disk
viscosity ($\nu(R) \propto R^{\gamma}$), and $\Sigma_t$ is the surface density 
at $R_t$ \citep{Isella2009}. The transition radius is the radius at which the 
mass flow is zero, such that for $R<R_t$ the mass flow goes inward and
mass is accreted into the disk, and for $R>R_t$ the mass flow goes outwards as 
the disk expands to conserve angular momentum. 

We assume that the central star has a bolometric luminosity of 
$L_{bol} = 30$~\lsun\ \citep{Cohen1983} and a mass of $M_* = 1$~\msun. 
The dust opacity, assumed to be constant throughout the disk, is calculated
using compact non-porous spherical grains with fractional abundances from
\citet{Pollack1994}: 12\% silicates, 27\% carbonaceous materials and 61\% ices.
The grain size distribution is assumed to be a power-law ($n(a) \propto
a^{-q}$), with slope $q=3.5$ and minimum grain size $a_{min} = 0.05$ $\mu$m. We
adopt a dust emissivity index of $\beta =1.23$ \citep[derived from a gray-body
fit to the spectral energy distribution;][]{Sandell1998}, and a dust-to-gas
ratio of 0.01. For these assumed parameters, the implied maximum grain size of
the distribution is $a_{max} = 0.1$~cm and the mass opacity corresponds to
$\kappa_{\nu} = 0.1$~cm$^{2}$ g$^{-1}$ at 1.3~mm.

The dust surface density defined in Equation~\ref{eq:sd} was fitted to the
observed visibilities using the procedure described in \citet{Isella2009}. 
Figure \ref{uvamp} compares the observed and modeled visibility profile for
PP~13S*. The model provides a reasonable fit to the data, although the 
observed visibility amplitudes are larger than the model for baselines longer 
than \about 1~km. A map of the model was created and subtracted from the 
observed PP~13S* map, but no significant residuals ($> 3\sigma$) were found. 

The best fit disk model has a disk inclination of 15\arcdeg\ (where 90\arcdeg\
is defined as an edge-on disk), $R_t=$13~AU, $\Sigma_t=$145~g~cm$^{-2}$,
$\gamma=0.95$, $M_\mathrm{disk}$=0.06~\msun, and $R_\mathrm{disk}$=128~AU. The
mass estimate is larger than the median mass of a Class II circumstellar disk
\citep{Andrews05}, and is in agreement with the measured masses around other
FU-Orionis objects. For example \citet{Sandell2001} estimated circumstellar
masses (disk and envelope) between 0.02~\msun\ to a few solar masses for a
sample of 16 FU-Orionis objects.

\begin{figure}
\begin{center}
\includegraphics[scale=0.7]{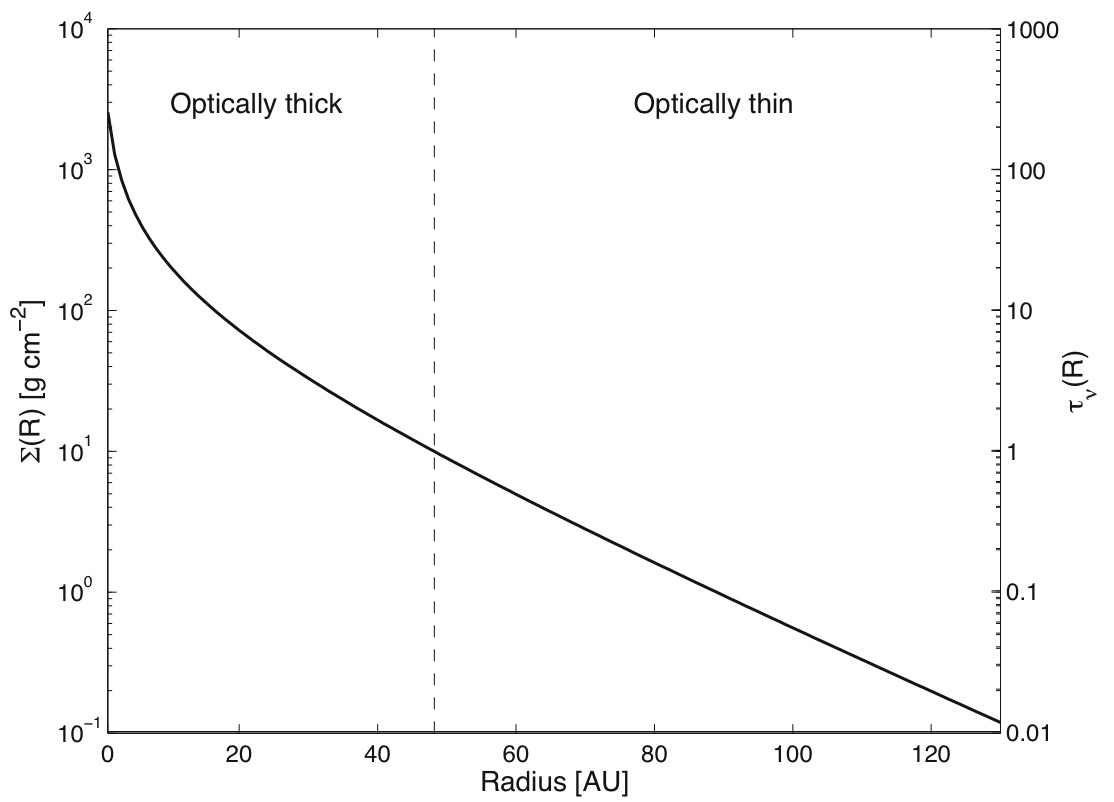} 
\caption{Model surface density distribution of PP~13S* as a function of 
disk radius.
The disk optical depth ($\tau = \Sigma_d(R) \times \kappa_{1.3mm}$) was 
computed assuming a constant dust opacity throughout the disk. The inner 
region of the disk ($R < 48$~AU) becomes optically thick at a frequency
of 227~GHz.} 
\label{surf_density}
\end{center} 
\end{figure}

Figure~\ref{surf_density} shows the surface density distribution
($\Sigma$) and the optical depth ($\tau$) as a function of the disk radius. The
vertical line marks the separation between the optically thin and thick
regimes. We find that the disk becomes optically thick inwards of 
$\sim 48$~AU given the assumptions in the model. We caution that the surface 
density distribution within this region is poorly constrained not only 
because of the high optical depth, but because the disk is unresolved 
for a radius less than 26~AU. The radius where the dust becomes optically 
thick is large compared to what is found in disks around classical T Tauri
stars, where only the inner few astronomical units are opaque at such long 
wavelength. Furthermore, the transition radius we find for PP~13S* is smaller
than any of the pre-main sequence circumstellar disks studied by
\citet{Isella2009}. Thus the circumstellar disk of PP~13S* is more
concentrated than the disks around classical T-Tauri stars.

Gravitational instabilities in the disk might be responsible for the
enhanced accretion episodes seen in FU-Orionis objects \citep{Armitage2001}. To
investigate if the disk around PP~13S* is gravitationally unstable, we
computed the Toomre's $Q$ parameter: 
\begin{eqnarray} 
   Q = \frac{c_s \kappa}{\pi G \Sigma(R)},
\end{eqnarray} 
where $c_s$ is the sound speed and $\kappa$ is the epicyclic frequency (which
is equal to the angular velocity for a keplerian disk). A disk becomes
gravitationally unstable if $Q \lesssim 1.5$, as spiral waves develop and 
mass is transported inwards and momentum is transported outwards
\citep{Lodato2004}. 

\begin{figure}
\begin{center}
\includegraphics[scale=0.7]{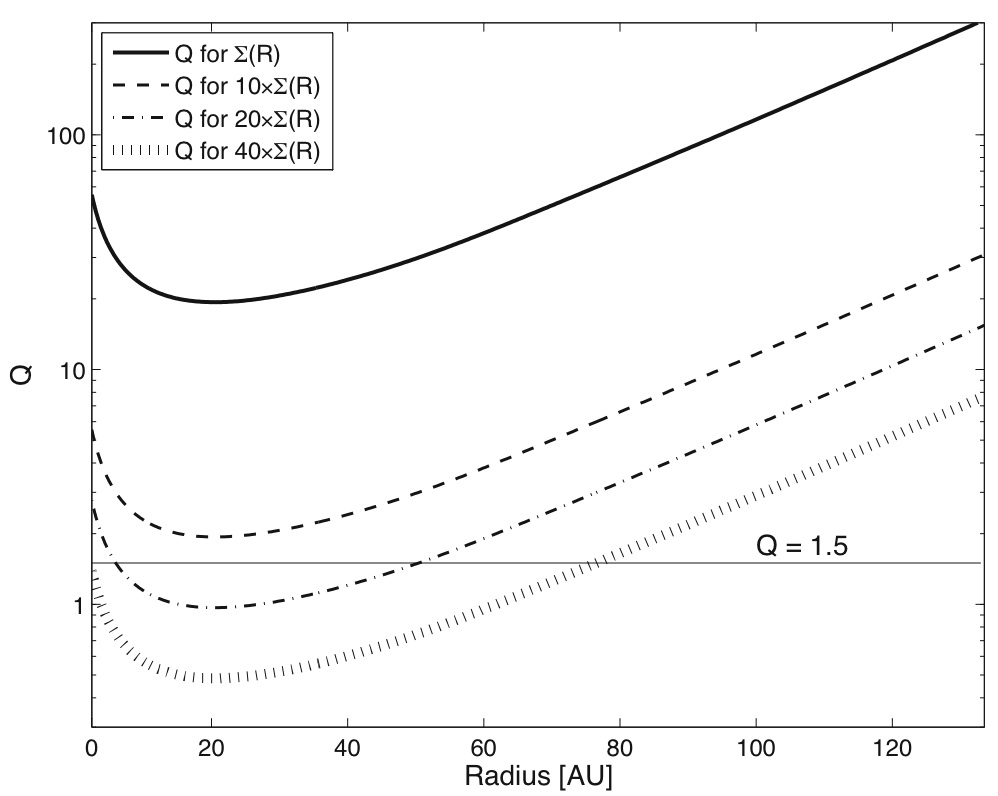} 
\caption{Toomre's instability parameter, $Q$, plotted for the PP~13S* disk. 
The solid line shows the $Q$ values for the nominal disk model. Dust opacities 
that are 10, 20 and 40 times larger than the adopted opacities are shown by the
dashed, dash-dotted, and vertical-dashed curves respectively. The $Q=1.5$ 
limit is shown for reference, such that regions with $Q\lesssim1.5$ may become 
gravitational unstable. The disk surface density needs to be an order of 
magnitude higher than the nominal model in order to form instabilities in 
the disk.
} 
\label{toomre} 
\end{center} 
\end{figure}

As shown in Figure~\ref{toomre}, the PP~13* disk is gravitationally stable
to axisymmetric perturbations across all radii for the inferred surface 
density. The disk surface density would need to
be increased by more than an order of magnitude (see dashed line of Figure
\ref{toomre}) for the disk to develop a gravitational instability. The inferred
disk surface density can vary widely depending on the adopted dust properties,
as composition, grain size distribution ($q$, $a_{min}$, $a_{max}$), emissivity
index ($\beta$) and dust-to-gas ratio, all affect the resulting opacity. For
example, $\kappa_{\nu}$ diminishes by 20-30\% if ices are ignored from the dust
composition, increasing the surface density by the same percentage. Also,
flattening the grain distribution slope to $q = 3.0$ gives a 30\% increase in
the disk mass. Furthermore, we adopted a fixed power-law ($\beta$) of the dust
opacity law. If $\beta$ decreases toward the center of the system \citep[as
observed in Class 0 sources;][]{Kwon2009} we can expect to have larger grains
in the circumstellar disk that will reduce the mass opacity and increase the
surface density. Despite the uncertainties in the dust properties, it will be
difficult to increase the surface density by more than an order of magnitude at
a radius $> 48$~AU where the disk is optically thin, unless the dust properties
of PP~13S* are extraordinarily different from what is found in typical disks
\citep{2007aAndrews, 2007bAndrews, Isella2009}. 

\section{Conclusions}

We have described C-PACS, which uses paired antennas as a means to
calibrate the atmospheric phase fluctuations on long interferometric 
baselines. Specifically, while the 6 and 10-m CARMA antennas observe a science
source in the 3-mm or 1-mm atmospheric windows, the 3.5-m CARMA antennas 
simultaneously observe a nearby atmospheric calibrator in the 1-cm band. The 
3.5-m antennas are placed within 30~m of the larger antennas to sample 
similar atmospheric delay fluctuations. We have applied the calibration 
technique to CARMA observations of the circumstellar material around the 
FU~Orionis object PP~13S$^*$. C-PACS yields quantitative improvement in the
image quality of PP~13S$^*$: the observed peak flux increased by a factor of
1.6, the image noise level decreased by 36\%, and the FWHM of the major axis decreased by 52\%.

The improvement in the phase error and amplitude coherence provided by the
paired antennas technique is a function of the projected beam separation of the
antenna pairs at the height of the turbulent layer, and the radiometric noise
introduced from the reference array to the science array. Thus, the brightness
of the atmospheric calibrator and the angular separation between the
atmospheric calibrator and the science target are the main restrictions to the
application of this technique for general science observations. Our current
estimate requires a calibrator closer than $5\degr$ to the science target and
brighter than 1~Jy at 30~GHz to correct 1.3~mm observations. Based on existing
radio catalogs, we estimate that there are 420 sources that have $S_{\nu} >
1$~Jy, such that 50\% of the sky can be observed with a suitable calibrator.

With C-PACS, we have obtained 0.15\arcsec\ resolution images of the
circumstellar material around PP~13S$^*$ at an observing frequency of 227~GHz.
We measure an integrated flux density of 241~mJy at 227~GHz, which is about
half of the extended emission detected in a $19.5''$ beam \citep{Sandell1998}.
We constrain the surface density profile of PP~13S* using a self-consistent
disk model. The main difference in the inferred disk properties compared to
disks around other pre-main sequence circumstellar disks is that the
dust is more centrally concentrated and there is a larger region that is 
optically thick at millimeter wavelengths. From analysis of the Toomre $Q$ 
parameter, we find that the disk is gravitationally stable over all disk radii 
unless the disk surface density is underestimated by an order of magnitude 
or more.

\acknowledgments

Acknowledgments: Support for CARMA construction was derived from the
Gordon and Betty Moore Foundation, the Kenneth T. and Eileen L. Norris
Foundation, the James S. McDonnell Foundation, the Associates of the California
Institute of Technology, the University of Chicago, the states of California,
Illinois, and Maryland, and the National Science Foundation. Ongoing CARMA
development and operations are supported by the National Science Foundation
under a cooperative agreement (grant AST 08-38260), and by the CARMA partner
universities. LMP acknowledges support for graduate studies through a
Fulbright-CONICYT scholarship. SM gratefully acknowledges support from an 
NSF Astronomy and Astrophysics Fellowship.
 
{\it Facilities:} \facility{CARMA}.

\end{document}